\documentclass[12pt]{article}
\usepackage[english,german,french,polish]{babel}
\usepackage[T1]{fontenc}
\usepackage{amsfonts}

\begin{document}

\selectlanguage{english}

\baselineskip 0.73cm
\topmargin -0.4in
\oddsidemargin -0.1in

\let\ni=\noindent

\renewcommand{\thefootnote}{\fnsymbol{footnote}}

\newcommand{\SM}{Standard Model }

\newcommand{\SMo}{Standard-Model }

\pagestyle {plain}

\setcounter{page}{1}

%\pagestyle{empty}

%\addtocounter{equation}{+1}

~~~~~~
\pagestyle{empty}

\begin{flushright}
IFT-- 09/9
\end{flushright}

\vspace{0.4cm}

{\large\centerline{\bf Is the electromagnetic field a source}}

{\large\centerline{\bf of a mediating field in hidden sector?}}

\vspace{0.5cm}

{\centerline {\sc Wojciech Kr\'{o}likowski}}

\vspace{0.3cm}

{\centerline {\it Institute of Theoretical Physics, University of Warsaw}}

{\centerline {\it Ho\.{z}a 69, 00--681 Warszawa, ~Poland}}

\vspace{0.6cm}

{\centerline{\bf Abstract}}

\vspace{0.2cm}
  
\begin{small}

%\begin{quotation}

We continue the discussion of a proposed model of hidden sector of the Universe, consisting of sterile spin-1/2
Dirac fermions ("\,$\!$sterinos"), sterile spin-0 bosons ("\,$\!$sterons") with spontaneously nonzero vacuum expectation 
value, and sterile nongauge mediating bosons ("\,$\!A$ bosons") described by an antisymmetric-tensor field (of dimension 
one).  The Standard-Model electromagnetic field (of dimension two) multiplied by the steron vacuum expectation value 
becomes in a spontaneous way a source of $A$-boson field mediating  new weak interactions in hidden sector and providing --- due to the action of electromagnetic field in both sectors --- a weak coupling between hidden and 
Standard-Model sectors ("photonic portal"). The proposed photonic portal emphasizes even more the role of electromagnetic field in the structure of the Universe.
 
\vspace{0.6cm}

\ni PACS numbers: 14.80.-j , 04.50.+h , 95.35.+d 

\vspace{0.6cm}

\ni September 2009

%\end{quotation}
 
\end{small}

\vfill\eject

\pagestyle {plain}

\setcounter{page}{1}

\vspace{0.5cm}

\ni {\bf 1. Introduction}

\vspace{0.4cm} 

In a recent work [1,2], we have conjectured the existence of a sterile nongauge antisymmetric-tensor field $A_{\mu \nu}$ (of dimension one) that is weakly coupled to the pair $\varphi F_{\mu\,\nu}$ of a sterile spin-0 boson field $\varphi$ and the \SMo electromagnetic field $F_{\mu\,\nu} = \partial_\mu A_\nu - \partial_\nu A_\mu$ (of dimension two), as well as to the pair $\bar{\psi}\psi$ of a spin-1/2 fermion field $\psi$ and its anifermion counterpart. So, such a conjecture leads to the new weak interaction

%rownanie 1%6
\begin{equation}
- \frac{1}{2} \sqrt{f}\left(\varphi F_{\mu \nu} + \zeta \bar\psi \sigma_{\mu \nu} \psi \right) A^{\mu \nu}\,,
\end{equation}

\ni where $\sqrt{f}$ and $\sqrt{\!f}\,\zeta$ denote two dimensionless small coupling constants ($f>0$). The interaction (1) appears in our model in addition to the familiar \SMo weak interaction (and also the very weak universal gravity).

Thus, in our model, the sterile field $A_{\mu \nu}$ mediates new weak interactions within a hypothetical hidden sector of the Universe (responsible for cold dark matter), consisting of sterile spin-1/2 fermions $\psi$ ("\,$\!$sterinos"), sterile spin-0 bosons $\varphi$ ("\,$\!$sterons") and sterile non-gauge spin-1 bosons $A_{\mu \nu}$ ("$A$ bosons") of two kinds with parity $-$ and +. It also provides a new weak coupling between the hidden and \SMo sectors due to the action of \SMo electromagnetic field $F_{\mu \nu}$ in both sectors (after the \SMo electroweak symmetry is spontaneously broken by the \SMo Higgs mechanism). We have called this second way of mediation the "photonic portal"\, to the hidden sector. Thus, our approach to the hidden sector differs from the popular one, where the "Higgs portal"\, to the hidden sector works [3].

In Refs. [1.2], we have assumed that

%rownanie 2%3
\begin{equation}
\varphi = <\!\!\varphi\!\!>_{\rm vac}\! + \,\varphi_{\rm ph}\,, 
\end{equation}

\ni where $<\!\!\varphi\!\!>_{\rm vac}$ is a spontaneously nonzero vacuum expectation value of the steron field $\varphi, $while $\varphi_{\rm  ph} \neq 0$ denotes  the physical steron field. The value $<\!\!\varphi\!\!>_{\rm vac} \neq 0$ has been used to generate the masses $m_\psi$, $m_\varphi$ and $M$ of sterinos, sterons and $A$ bosons. The mass scale $M$ of $A$ bosons is typically expected to be large, but a moderate value for it is not {\it a priori} excluded. In addition, the term with $<\!\!\varphi\!\!>_{\rm vac} \neq 0$ in the interaction (1) generates spontaneously a tiny sterino magnetic moment

%rownanie 3%10
\begin{equation}
 \mu_\psi = \frac{f \zeta}{2M^2}<\!\!\varphi\!\!>_{\rm vac} \,,
\end{equation}

\ni since the coupling (1) implies for sterinos (though they are electrically neutral) the effective magnetic interaction

%rownanie 4%9
\begin{equation}
- \frac{f \zeta}{2M^2} <\!\!\varphi\!\!>_{\rm vac} \bar\psi \sigma_{\mu \nu} \psi F^{\mu \nu}  
\end{equation}

\ni ({\it cf.} the effective coupling (8)).

Note that with the use of interaction Lagrangian (1) the following field equations hold for $A_{\mu \nu}$ and $F_{\mu \nu}$: 

%rownanie 5%1
\begin{equation}
(\Box - M^2)A_{\mu \nu} = - \sqrt{f} (\varphi F_{\mu \nu} + \zeta \bar\psi \sigma_{\mu \nu} \psi) 
\end{equation}

\ni  and

%rownanie 6%2
\begin{equation}
\partial^\nu (F_{\mu \nu} +  \sqrt{\!f}\, \varphi A_{\mu \nu}) = -j_\mu \;\;,\;\; F_{\mu \nu} = \partial_\mu A_\nu - \partial_\nu A_\mu 
\end{equation}

\ni ("$\!$\,supplemented Maxwell's equations"). Here, $j_{\mu}$ is the familiar \SMo electric current.

The field equation (5) shows that, beside the pair $\bar\psi \sigma_{\mu\,\nu} \psi$, the field pair $\varphi F_{\mu \nu}$ is a source of the massive field $A_{\mu \nu}$. Due to Eq. (2) with $<\!\!\varphi\!\!>_{\rm vac} \neq 0$, the pair $\varphi F_{\mu \nu}$ in Eq. (5) generates spontaneously a source term $-\sqrt{\!f\,} \!\!<\!\!\varphi\!\!>_{\rm vac}\!F_{\mu \nu}$ for the sterile mediating field $A_{\mu \nu}$, which is provided by the electromagnetic field $F_{\mu \nu}$ alone.

If momentum transfers mediated through the field $A_{\mu \nu}$ in virtual states are negligible {\it versus} the mass scale $M$, then it follows from Eq. (5) that approximately 

%rownanie 7%7
\begin{equation}
 A_{\mu \nu} \simeq \frac{\sqrt{f}}{M^2}\left(\varphi F_{\mu \nu} + \zeta \bar\psi \sigma_{\mu \nu} \psi \right)\,.
\end{equation}

\ni Then, the interaction (1) gives approximately the effective coupling (in an effective Lagrangian)

%rownanie 8%8
\begin{equation}
  -\frac{1}{4} \frac{f}{M^2}\left(\varphi F_{\mu \nu} + \zeta \bar\psi \sigma_{\mu \nu} \psi \right) \left(\varphi F^{\mu \nu} + \zeta \bar\psi \sigma^{\mu \nu} \psi \right) 
\end{equation}

\ni being an analogy of the Fermi coupling. Here, $\varphi =<\!\!\varphi\!\!>_{\rm vac}\! + \,\varphi_{\rm ph}$.

In Ref. [2], the new weak interaction (1) has been conjectured (more correctly) to be embedded in a more extended weak interaction displaying the overall electroweak symmetry spontaneously broken by the \SMo Higgs mechanism. In the present note, we will restrict ourselves to the simpler interaction (1).

\vspace{0.3cm}  

{\bf 2. Two kinds of sterile $A$ bosons}

\vspace{0.3cm}

The antisymmetric-tensor field $A_{\mu \nu}$ can be split into a vector and an axial three-dimensional fields $\vec{A}^{\rm (E)} = \left(A^{\rm (E)}_k \right)$ and $\vec{A}^{\rm (B)} = \left(A^{\rm (B)}_k \right)\;(k = 1,2,3)$  of spin 1 and parity $-$ and +, respectively, in an analogy to the splitting of  the electromagnetic field $F_{\mu \nu} = \partial_\mu A_\nu - \partial_\nu A_\mu $ into the electric and magnetic three-dimensional fields $\vec{E} =\left(E_k\right)$ and $\vec{B} =\left(B_k\right)\;(k = 1,2,3)$,

%rownanie 9%15
\begin{equation} 
\left(F_{\mu \nu}\right) = \left(\begin{array}{rrrr} 0 & E_1 & E_2 & E_3 \\ -E_1 & 0\;\; & -B_3 & B_2 \\ -E_2 & B_3 & 0\;\; & -B_1 \\  -E_3 & -B_2 & B_1 & 0\;\; \end{array} \right) \,.
\end{equation}

\ni Here, 

%rownanie 10%16  
\begin{equation} 
\vec{E} = -\partial_0 \vec{A} - \vec{\partial} A_0 = (-F_{k  0}) \;\;\;,\;\;\; \vec{B}  = \vec{\partial}\times \vec{A} = \left(- \frac{1}{2}\varepsilon_{k l m} F_{l m}\right)
\end{equation}

\ni with $\vec{A} = (A^k) = (-A_k)$ and $\vec{\partial} = (\partial_k) - (-\partial^k)\;(k=1,2,3)$. In fact, defining 

%rownanie 11
\begin{equation}
\vec{A}^{(E)} = \left(- A_{k 0}\right) \;\;\; \;,\;\;\; \vec{A}^{(B)} = \left(- \frac{1}{2}\varepsilon_{k l m} A_{l m}\right)
\end{equation}%w194

\ni $(k = 1,2,3)$, we obtain

%rownanie 12%14
\begin{equation} 
\left(A_{\mu \nu}\right) = \left(\begin{array}{rrrr} 0\;\;\;\;  & A^{(E)}_1 & A^{(E)}_2 & A^{(E)}_3 \\ -A^{(E)}_1 & 0\;\; \;\; & -A^{(B)}_3 & A^{(B)}_2 \\ -A^{(E)}_2 & A^{(B)}_3 & 0\;\; \;\; & -A^{(B)}_1 \\  -A^{(E)}_3 & -A^{(B)}_2 & A^{(B)}_1 & 0\;\; \;\; \end{array} \right)  = \left(A^{(E)}_{\mu \nu}\right) + \left(A^{(B)}_{\mu \nu}\right) \,. 
\end{equation}

From Eqs. (9) and (12) it follows that

%rownanie 13%18
\begin{equation} 
F^{\mu \nu} F_{\mu \nu} = - 2\left(\vec{E}^2 - \vec{B}^2\right)\;\;\;,\;\;\;A^{\mu \nu} A_{\mu \nu} =  -2\left( \vec{A}^{(E)\,2} - \vec{A}^{(B)\,2}\right) 
\end{equation}

\ni and

%rownanie 14%17
\begin{equation} 
F^{\mu \nu} A_{\mu \nu} = - 2\left(\vec{E}\cdot \vec{A}^{(E)} - \vec{B}\cdot \vec{A}^{(B)}\right) \,.
\end{equation}

On the other hand, making use of the matrix

%rownanie 15%19
\begin{equation} 
(\sigma^{\mu \nu}) = \left(\begin{array}{rrrr} 0\;\;  &i \alpha_1 & i \alpha_2 & i \alpha_3 \\ -i \alpha_1 & 0\;\; & \sigma_3 & -\sigma_2 \\ -i \alpha_2 & -\sigma_3 & 0\;\; & \sigma_1 \\  -i \alpha_3 & \sigma_2 & -\sigma_1 & 0\;\; \end{array} \right) 
\end{equation}

\ni of spin tensor $\sigma^{\mu \nu} = (i/2)[\gamma^\mu,\gamma^\nu]$, where $\vec{\alpha} = (\alpha_k) = (\gamma^0\gamma^k) = (-i\sigma_{k 0})$ and $\vec{\sigma}=(\sigma_k) = (\gamma_5\vec{\alpha}) = \left((1/2)\varepsilon_{k l m} \sigma_{l m}\right) \;(k=1,2,3)$, we get

%rownanie 16%20
\begin{equation} 
\sigma^{\mu \nu}A_{\mu \nu} = 2\left(i \vec{\alpha}\cdot \vec{A}^{\rm (E)} - \vec{\sigma}\cdot \vec{A}^{\rm (B)}\right)\,.
\end{equation}

In consequence of Eqs. (14) and (16), we can rewrite the interaction (1) as

%rownanie 17%21
\begin{equation}
\sqrt{f}\left[\left(\varphi \vec{E}- i\zeta \bar\psi \,\vec{\alpha} \,\psi \right)\cdot \vec{A}^{E}- \left(\varphi \vec{B} - \zeta \bar{\psi} \,\vec{\sigma}\psi \right)\!\cdot\!\vec{A}^{(B)}\right] 
\end{equation}%249

\ni and so, the field equation [5] for $A_{\mu \nu}$ in the form

%rownanie 18%22
\begin{eqnarray} 
(\Box - M^2)\vec{A}^{(E)} & = & - \sqrt{f}\left(\varphi \vec{E} - i \zeta \bar{\psi}\,\vec{\alpha}\,\psi \right)\,, \nonumber \\ (\Box - M^2)\vec{A}^{(B)} & = & - \sqrt{f}\left(\varphi \vec{B}\, - \,\zeta\bar{\psi}\, \vec{\sigma}\,\psi \right)\,.
\end{eqnarray}

\ni Here, as usual, $\varphi = <\!\!\varphi\!\!>_{vac}\! + \,\varphi_{\rm ph}$ with $<\!\!\varphi_{\rm vac}\!\!> \neq 0$.

We can see from Eqs. (18) that the electric and magnetic fields $\vec{E}$ and $\vec{B}$ are involved respectively in the sources of two different sterile mediating fields $\vec{A}^{\rm (E)}$ and $\vec{A}^{\rm (B)}$ being parts of the relativistic structure $A_{\mu \nu}$. In particular, in Eqs (18) there are source terms
 
%rownanie 19%23
\begin{equation} 
-\sqrt{f}<\!\!\varphi\!\!>_{\rm vac}\vec{E} \;\; \,,\,\;\; -\sqrt{f}<\!\!\varphi\!\!>_{\rm vac}\vec{B}
\end{equation}

\ni spontaneously generated by $<\!\!\varphi\!\!>_{\rm vac}$ and provided by $\vec{E}$ and $\vec{B}$ fields alone.

The sterile $A$ bosons of two kinds described by the fields $\vec{A}^{ (E)}$ and $\vec{A}^{(B)}$ --- when they  propagate freely in space --- get the following wave functions: 

%rownanie 20%25
\begin{equation} 
\vec{A}^{\,(E,B)}_{\vec{k}}(x) = \frac{1}{(2\pi)^{3/2}} \frac{1}{\sqrt{2 \omega_A}}\, \vec{e}^{\;(E,B)} e^{-i k_A\cdot x } \,,
\end{equation}

\ni where $k_A = (\omega_A, \vec{k}_A)$ with $\omega_A =\sqrt{\vec{k}_A^2 + M^2}$, while $\vec{e}^{\;(E,B)} = \vec{e}_a^{\;(E,B)}\;\;(a=1,2,3)$ are three orthonormal linear polarizations for $A^{(E)}$ and $A^{(B)}$ bosons,  satisfying the formulae

%rownanie 21%26
\begin{equation}
\vec{e}_a^{\;(E,B)}\cdot \vec{e}_b^{\;(E,B)} = \delta_{a b}\;\; (a,b = 1,2,3) \;\;\;,\;\;\; \sum^3_{a=1} {e}_{a k}^{\,(E,B)} {e}_{a l}^{\,(E,B)} = \delta_{k l}
\end{equation}

\ni with $\vec{e}_a ^{\;(E,B)} = ({e}_{a k}^{\;(E,B)}) \; (a=1,2,3)$.

\vspace{0.3cm}

\ni {\bf 3. Decays of $A^{\,(E,B)}$ bosons into charged-fermion pairs} 

\vspace{0.3cm}

The sterile mediating bosons ${A}^{ (E,B)}$ are unstable. Their simple decays are those into pairs of charged fermions, 
 $A^{(E,B)} \rightarrow \gamma^* \rightarrow \bar{f} f$, if $M > 2 m_f$ (here, for instance, $f = e^-$ or $p$; if $M<2 m_p$, the decays $A^{(E,B)} \rightarrow \bar{p} p$ are forbidden). Such processes are described by the coupling

%rownanie 22%35
\begin{equation}
\sqrt{f}\,<\!\!\varphi\!\!>_{\rm vac} \left(\vec{E}\cdot \vec{A}^{(E)} - \vec{B}\cdot \vec{A}^{(B)}\right) 
\end{equation}

\ni being a part of the interaction (17), jointly with the Standard-Model electromagnetic interaction $-e_f \bar{\psi}_f  \gamma^\mu \psi_f A_\mu$ of $f$ fermions. In this case, the corresponding $S$ matrix elements take in the lowest order the form (in an obvious notation) [1]:

%rownanie 23%36
\begin{eqnarray}
S\!\left(A^{(E)} \!\rightarrow\! \bar{f} f \right)\!\!\! & = & \!\!\!e_{\!f} \sqrt{f} <\!\!\varphi\!\!>_{\rm vac}\left[\frac{1}{(2\pi)^9}\frac{m^2_f}{E_1 E_2 2\omega_{A}}\right]^{1/2}\!\!\left[\bar{u}_f(p_1)\frac{1}{i}\left(\omega_{A}\vec{\gamma} \!-\! \vec{k}_{A}\beta\right)\!\cdot\!\vec{e}^{\,(E)}v_f(p_2)\right] \nonumber \!\!\! \\ & & \!\!\! \times \frac{1}{k^2_A} (2\pi)^4 \delta^4(p_1 +p_2 - k_A)
\end{eqnarray}

\ni and

%rownanie 24%37
\begin{eqnarray}
S\!\left(A^{(B)}\rightarrow \bar{f} f \right)\!\!\! & = & e_{\!f} \sqrt{f} <\!\!\varphi\!\!>_{\rm vac}\left[\frac{1}{(2\pi)^9}\frac{m^2_f}{E_1 E_2 2\omega_{A}}\right]^{1/2} \!\!\left[\bar{u}_f(p_1)\frac{1}{i}\left(\vec{k}_A\times\vec{\gamma}\right)\cdot\vec{e}^{\,(B)} v_f(p_2) \right]  \nonumber \\  & & \!\!\! \times \frac{1}{k^2_A} (2\pi)^4 \delta^4(p_1 +p_2 - k_A)\,.
\end{eqnarray}

Then, the differential decay rates

%rownanie 25%38
\begin{eqnarray}
\frac{d^{\,6}\Gamma\!\left(A^{ (E,B)} \rightarrow \bar{f} f\right)}{d^3\vec{p}_1 d^3\vec{p}_2} = (2\pi)^3\sum_{u\,v}\frac{1}{3} \sum_{e^{ (E,B)}} \frac{|S\!\left(A^{ (E,B)} \rightarrow \bar{f} f\right)|^2}{(2\pi)^4 \delta^4(0)} 
\end{eqnarray}

\ni give

%rownanie 26%39
\begin{eqnarray}
\frac{d^{\,6}\Gamma\!\left(A^{ (E)} \rightarrow \bar{f} f\right)}{d^3\vec{p}_1 d^3\vec{p}_2} & = & \frac{e^2_f f<\!\varphi\!\!>^2_{\rm vac}}{(2\pi)^2} \frac{1}{6E_1 E_{2\,}\omega_{\!A}} \left[1 - 2\frac{E_1 E_2}{M^2} + 2\left(\frac{m_f\omega_A}{M^2}\right)^{\!\!2}\right] \nonumber \\ & & \times\delta^4(p_1 + p_2 - k_A) 
\end{eqnarray}
\ni and

%rownanie 27%40
\begin{eqnarray}
\frac{d^{\,6}\Gamma\!\left(A^{(B)} \rightarrow \bar{f} f\right) }{d^3\vec{p}_1 d^3\vec{p}_2} & = & \!\!\frac{e^2_f f<\!\!\varphi\!\!>^2_{\rm vac}}{(2\pi)^2}  \frac{1}{6E_1 E_{2\,}\omega_{\!A}} \left[\frac{\vec{k}^2_A}{M^2} - \left(\frac{\vec{p}_1\times \vec{k}_A}{M^2}\right)^{\!\!2} \!-\! \left(\frac{\vec{p}_2\times \vec{k}_A}{M^2}\right)^{\!\!2} \!\right] \nonumber \\ & & \times \delta^4(p_1 + p_2 - k_A) \,.
\end{eqnarray}

At rest (when $\vec{k}_A =0$), the total decay rates in the channels $A^{ (E,B)} \rightarrow \bar{f} f$, 

%rownanie 28%42
\begin{equation}
\Gamma\!\left(A^{ (E,B)} \rightarrow \bar{f} f\right) = \int d^3\vec{p}_{1}\,d^3\vec{p}_2 \frac{d^{\,6}\Gamma\!\left(A^{ (E,B)} \rightarrow \bar{f} f\right)}{d^3\vec{p}_{1}\,d^3\vec{p}_2} \,,
\end{equation}

\ni are

%rownanie 29%42
\begin{equation}
\Gamma\!\left(A^{ (E)} \rightarrow \bar{f} f\right) = \frac{e^2_f\, f \!<\!\!\varphi\!\!>^2_{\rm vac}}{24\pi M}\left[1 + \left(\frac{2m_f}{M}\right)^{\!\!2}\right] \left[1 - \left(\frac{2m_f}{M}\right)^{\!\!2}\right]^{\!1/2}  
\end{equation}

\ni and

%rownanie 30
\begin{equation}
\Gamma\!\left(A^{ (B)} \rightarrow \bar{f} f\right) = 0\,,
\end{equation}

\ni where $\vec{p}_1 + \vec{p}_2 = \vec{k}_A = 0$ and $E_1 + E_2 = \omega_A = M$ (so, $E_1 = E_2 =M/2$ and $ |\vec{p}_1| = |\vec{p}_2| = \sqrt{(M/2)^2 - m^2_f\,}\,$ with $M > 2m_{\!f} $). 

If tentatively $f \sim e^2$, $\zeta \sim 1$ and $m^2_\psi\sim m^2_\varphi(1\;{\rm to}\; 1/4) \sim \,<\!\!\varphi\!\!>^2_{\rm vac}\, \sim M^2 (1 \;{\rm to}\; 10^{-4})$, then it follows from the thermal condition (39) for sterinos discussed in Section 4 that $M \sim (650 \;{\rm to}\; 1.6)$ GeV, and the formula (29) gives

%rownanie 31
\begin{equation}
\Gamma\!\left(A^{ (E)} \rightarrow \bar{f} f\right) \sim \frac{e^4 M(1 \;{\rm to}\, 10^{-4})}{24 \pi}\sim 72\,{\rm MeV \; to}\; 17\,{\rm eV} = 1.1\times 10^{23}\, \frac{1}{\rm s}\;{\rm to}\; 2.6\times 10^{16}\, \frac{1}{\rm s}\,,  
\end{equation}

\ni when $e^2_f = e^2$ and $m^2_f \ll M^2 \;(\hbar = 1 = c)$.

A simple production process for sterile mediating bosons $A^{ (E,B)}$ is the inelastic Compton effect $\gamma f \rightarrow \gamma^* f \rightarrow A f$ at energies high enough to produce the mass $M$.
 
\vspace{0.3cm}

\ni {\bf 4. Sterinos as candidates for thermal dark matter} 

\vspace{0.3cm}

In our model of hidden sector of the Universe, stable sterinos are natural candidates for the thermal cold dark matter 
(sterons are unstable, decaying {\it e.g.} as $\varphi_{\rm ph} \rightarrow \gamma A^* \rightarrow \gamma \gamma$ through the part $-(1/2)\sqrt{f}\left(<\!\!\varphi\!\!>_{\rm vac}\! + \,\varphi_{\rm ph}\right) F_{\mu \nu} A^{\mu \nu}$ of the interaction (1)).

The abundance of dark matter presently observed by WMAP, $\Omega_{\rm DM} h^2 \simeq 0.11$, gives for the thermal average of total annihilation cross-section of a weakly interacting sterino-antisterino pair (multiplied by sterino relative velocity) the familiar thermodynamical estimate [4,1]

%rownanie 32
\begin{equation}
<\!\sigma_{\rm ann}\, v_{\rm DM}\!> \,\simeq\, {\rm pb} \simeq \frac{8}{\pi} \frac{10^{-3}}{{\rm TeV}^2}
\end{equation}

\ni (pb = $10^{-36}$ cm$^2$, $c = 1 = \hbar$). We will put here

%rownanie 33
\begin{equation}
\sigma_{\rm ann} \simeq\, \sigma(\bar{\psi} \psi \rightarrow e^+ e^-) + \sigma(\bar{\psi} \psi \rightarrow \varphi_{\rm ph}\gamma)\,,
\end{equation}

\ni where the processes $\bar{\psi} \psi \rightarrow \gamma^* \rightarrow e^+ e^-$ and $\bar{\psi} \psi \rightarrow  \varphi_{\rm ph}\gamma$ are approximately described by the parts of effective coupling (8),

%rownanie 34
\begin{equation}
-\frac{1}{2}\; \frac{f \zeta}{M^2}{<\!\!\varphi\!\!>_{\rm vac}} F_{\mu \nu} \bar\psi \sigma^{\mu \nu} \psi \,,
\end{equation}

\ni jointly with the electron electromagnetic interaction $e\, \bar\psi_e \gamma^{\mu} \psi_e A_\mu$, and

%rownanie 35
\begin{equation}
-\frac{1}{2}\; \frac{f \zeta}{M^2} \,\varphi_{\rm ph}\, F_{\mu \nu} \bar\psi \sigma^{\mu \nu} \psi \,, 
\end{equation}

\ni respectively.

Then, we  calculate in the sterino-antisterino centre-of-mass frame (where $v_{\rm DM} = 2v_\psi $) the following annihilation cross-sections [1]:

%rownanie 36
\begin{equation}
\sigma(\bar{\psi} \psi \rightarrow e^+ e^-) 2v_\psi = \frac{1}{12\pi}  \left(\frac{e\, f \zeta\!<\!\!\varphi\!\!>_{\rm vac}}{M^2}\right)^{\!\!2}\left(1+\frac{2m^2_\psi}{E^2_\psi}\right) 
\end{equation}

\ni  and

%rownanie 37
\begin{equation}
\sigma(\bar{\psi} \psi \rightarrow \varphi_{\rm ph} \gamma) 2v_\psi = \frac{1}{6\pi} 
\left(\frac{f \zeta}{M^2}\right)^{\!\!2} \left(1+ \frac{2m^2_\psi}{E^2_\psi}\right) \left(E^2_\psi -   \frac{m^2_\varphi}{4}\right) \,,
\end{equation}

\ni where in the first channel the electron mass $m_e$ is neglected ($E_e = E_\psi \geq m_\psi \gg m_e$). Hence, as $ E_\psi \simeq m_\psi$ for cold dark matter, we get

%rownanie 38
\begin{equation}
\sigma_{\rm ann} v_{\rm {DM}} \simeq \!\left[\sigma(\bar{\psi} \psi \rightarrow e^+ e^-) + \sigma(\bar{\psi} \psi \rightarrow \varphi_{\rm ph} \gamma)\!\right] 2v_\psi = \frac{1}{2\pi} \!\left(\frac{f \zeta}{M^2}\right)^{\!\!2}\!\!\left(\frac{e^2}{2}\!<\!\!\varphi\!\!>^2_{\rm vac} \!+\, m^2_\psi \!-\! \frac{m^2_\varphi}{4}\right)
\end{equation}

\ni ($e^2 = 4\pi \alpha = 0.0917$ with $\alpha = 1/137$). Here, $<\!\sigma_{\rm ann}\, v_{\rm DM}\!>  =\sigma_{\rm ann}\, v_{\rm DM}$. Comparing Eqs. (32) and (38), we obtain

%rownanie 39
\begin{equation}
\frac{1}{16}\left(f \zeta\right)^{\!2}\!\! \left(\frac{e^2}{2} \!<\!\!\varphi\!\!>^2_{\rm vac}\! +\, m^2_\psi - \frac{m^2_\varphi}{4}\right) \simeq  \frac{M^4}{{\rm TeV}^2}\times 10^{-3} \,.
\end{equation}

\ni This is a condition for sterinos to be thermal cold dark matter in our model of hidden sector.

Putting tentatively $f \sim e^2\,,\, \zeta \sim 1$ and $m^2_\psi \sim m^2_\varphi \sim \,<\!\!\varphi\!\!>^2_{\rm vac}\, \sim M^2$, we estimate from the condition (39) that $M \sim 650$ GeV. Alternatively, putting tentatively $f \sim e^2\;,\; \zeta \sim 1$ and $m^2_\psi \sim m^2_\varphi/4 \sim \,<\!\!\varphi\!\!>^2_{\rm vac}\, \sim M^2$, we calculate that $M \sim 160$ GeV. Eventually, if tentatively  $f \sim e^2\;,\; \zeta \sim 1$ and $m^2_\psi \sim m^2_\varphi/4 \sim \,<\!\!\varphi\!\!>^2_{\rm vac}\,\sim M^2\times 10^{-4}$, then $M \sim 1.6$ GeV (here, $M$ is $ < 2 m_p$). In the last case, particles of the hidden sector are light ($m_\psi \sim m_\varphi/2 \sim 16$ MeV). 

\vspace{0.3cm}

\ni {\bf 5. Conclusions} 

\vspace{0.3cm}

In this note, we have emphasized a two-level structure of the proposed model of hidden sector of the Universe weakly coupled with its \SMo sector through the "photonic portal". On the first level, the \SMo electric current $j_\mu$ acts as the source of the electromagnetic gauge field $F_{\mu \nu}$, 

%rownanie 40
\begin{equation}
\partial^{\,\nu}\!\left[F_{\mu \nu} +  \sqrt{f} (<\!\!\varphi\!\!>_{\rm vac}\! + \,\varphi_{\rm ph}) A_{\mu \nu}\right] = -j_\mu \;\;\;\;,\;\;\; F_{\mu \nu} = \partial_\mu A_\nu - \partial_\nu A_\mu\,, 
\end{equation}

\ni while on the second level, the electromagnetic field $F_{\mu \nu}$ multiplied by $<\!\!\varphi\!\!>_{\rm vac} \neq 0$ becomes in a spontaneous way a source of a sterile nongauge field $A_{\mu \nu}$ mediating interactions in the hidden sector,

%rownanie 41
\begin{equation}
(\Box - M^2)A_{\mu \nu} = - \sqrt{f} \left[(<\!\!\varphi\!\!>_{\rm vac}\! + \,\varphi_{\rm ph})F_{\mu \nu} + \zeta \bar\psi \sigma_{\mu \nu} \psi\right] \,. 
\end{equation}

\ni A simple two-level process is here $A \rightarrow \gamma^* \rightarrow \bar{f} f$ described in Section 3. The familiar \SM has, on the contrary, a one-level structure, where its currents, treated on the same footing, are sources of all their gauge fields.

Beside Eqs. (40) and (41), the sterino and steron fields, $\psi$ and $\varphi = <\!\!\varphi\!\!>_{\rm vac}\!+ \,\varphi_{\rm ph}$, satisfy the sterile-matter field equations 

%rownanie 42
\begin{equation}
\left(i \gamma^\mu \partial_\mu - \frac{1}{2}\sqrt{f}\,\zeta \sigma_{\mu \nu} A^{\mu \nu} - m_\psi\right) \psi = 0 
\end{equation}

\ni and

%rownanie 43
\begin{equation}
(\Box - m^2_\varphi) \varphi_{\rm ph} = \frac{1}{2}\sqrt{f} \, F_{\mu \nu} A^{\mu \nu} \;. 
\end{equation}

The masses of sterile particles, $m_\psi$, $m_\varphi$ and $M$, can be spontaneously generated by $<\!\!\varphi\!\!>_{\rm vac} \neq 0$ (what introduces necessarily additional weak couplings of $\varphi_{\rm ph}$ in Eqs. (40)---(43)) [1].

As mentioned at the end of Introduction, our proposed two-level structure of the Universe can be embedded in a more extended two-level structure displaying the overall electroweak symmetry spontaneously broken by the \SMo Higgs mechanism [2].

\vfill\eject

\vspace{0.4cm}

{\centerline{\bf References}}

\vspace{0.4cm}

\baselineskip 0.73cm

{\everypar={\hangindent=0.65truecm}
\parindent=0pt\frenchspacing

{\everypar={\hangindent=0.65truecm}
\parindent=0pt\frenchspacing

~[1]~W.~Kr\'{o}likowski, {\it Acta Phys. Polon.} {\bf B 39}, 1881 (2008); arXiv: 0803.2977 [{\tt hep--ph}]; {\it Acta Phys. Polon.} {\bf B 40}, 111 (2009); arXiv: 0903.5163 [{\tt hep--ph}].

\vspace{0.2cm}

~[2]~W.~Kr\'{o}likowski, arXiv: 0905.3987 [{\tt hep--ph}].

\vspace{0.2cm}

~[3]~{\it Cf. e.g.}  J. March-Russell, S.M. West, D. Cumberbath and D.~Hooper, {\it J. High Energy Phys.} {\bf 0807}, 058 (2008); K.~Kohri, J.~McDonald and N.~Sahu, arXiv: 0905.1312 [{\tt hep-ph}]; and references therein.

\vspace{0.2cm}

~[4]~{\it Cf. e.g.} E.W.~Kolb and S.~Turner, {\it Early Universe}, Addison-Wesley, Reading, Mass., 1994; K.~Griest and D.~Seckel, {\it Phys. Rev.} {\bf D 43}, 3191 (1991); G.~Bartone, D.~Hooper and J.~Silk, {\it Phys. Rep.} {\bf 405}, 279 (2005); M.~Taoso, G.~Bartone and A.~Masiero, arXiv: 0711.4996 [{\tt astro-ph}]; S.~Weinberg, {\it Cosmology}, Oxford University Press, New York 2008. 

\vspace{0.2cm}

\vfill\eject

\end{document}